\documentclass{iau}
\usepackage{graphicx}

\title[X-raying circumstellar disks] 
{X-raying circumstellar material around young stars}

\author[Schneider \& G\"unther]   
{P. C. Schneider$^1$
 \and H. M. G\"unther$^2$}

\affiliation{$^1$ESA/ESTEC\\email: {\tt cschneid@cosmos.esa.int}\\
$^2$MIT\\email: {\tt hgunther@mit.edu}}

\pubyear{2015}
\volume{314}  
\pagerange{119--126}
\jname{Young Stars \& Planets Near the Sun}
\editors{J. H. Kastner, B. Stelzer, \& S. A. Metchev, eds.}
\begin{document}

\maketitle

\begin{abstract}
Young stars are surrounded by copious amounts of circumstellar material. Its composition, in particular
its gas-to-dust ratio, is an important parameter. However, measuring this ratio is 
challenging, because gas mass estimates are often model dependent. 
X-ray absorption is sensitive to the gas along the line-of-sight while optical/near-IR 
extinction depends on the dust content. 
Therefore, the gas-to-dust ratio of an absorber is given by the ratio between X-ray and 
optical/near-IR extinction. We present three systems where
we used X-ray and optical/near-IR data to constrain the gas-to-dust 
ratio of circumstellar material; from a dust-rich debris disk to gaseous protoplanetary disks.

\end{abstract}

\firstsection 
\section{Introduction}
The chemical composition of circumstellar disks is one of their most important properties
as it controls many key phenomena like accretion, angular momentum transport, and 
ionization. Therefore, it  directly impacts planet formation
in protoplanetary disks. A basic parameter describing the chemical
composition is the gas-to-dust ratio. The interstellar medium (ISM) consists mainly 
of gas; only a small fraction ($\sim1$\,\%) is
in the form of dust grains. Protoplanetary disks are thought to have a gas-to-dust ratio
close to the ISM value 
during their early evolutionary stages. Therefore, the ISM gas-to-dust ratio of 100:1
is often \emph{assmued} when deriving the total disk mass from dust thermal emission 
measurements.
Directly deriving the total gas reservoir is more challenging
as the conversion of  gas emission line fluxes to total (gas) disk mass strongly depends on
often uncertain model parameters. 

One possibility to trace the gas is X-ray absorption. The absorption cross section is 
essentially independent of the chemical bonds of the absorbing
material, because X-ray absorption involves the inner shell electrons while 
molecular bonds involve the outer (valence) electrons. Thus, we can derive the absorber's 
gas content from its X-ray absorption signal.
Comparison with optical extinction caused by the dust along the line-of-sight
enables us to measure the gas-to-dust ratio of the absorbing material.

The location of the absorber(s) along the line-of-sight often remains
uncertain. Time variability of the absorption can help to overcome this ambiguity. 
In this proceedings, we  present three cases where the absorber location
can be determined with some certainty so that a local gas-to-dust ratio can be 
derived.

\vspace*{-0.3cm}
\section{The debris disk around AU Mic}
AU~Mic is a young ($23 \pm 3$ \,Myrs, \cite{Mamajek_2015}) M1 dwarf at a distance of about 10\,pc belonging to the $\beta\,$Pic
moving group (\cite{Zuckerman_2001}). 
AU~Mic is surrounded by an edge-on debris disk, i.e., a dust-rich disk. 
This disk, extending out to \hbox{$\sim150\,$AU},  has been subject to numerous 
studies from scattered light (e.g. \cite{Kalas_2004, Krist_2005}), over dust emission (\cite{Liu_2004})
to absorption (\cite{Roberge_2005}) and molecular hydrogen emission (\cite{France_2007}).
Given its small distance to the Earth, we can exclude substantial interstellar absorption  so 
that any absorption signal can be safely associated with AU~Mic's
circumstellar disk.

\begin{figure}[t]
\begin{center}
\includegraphics[width=2.6in,height=2.1in]{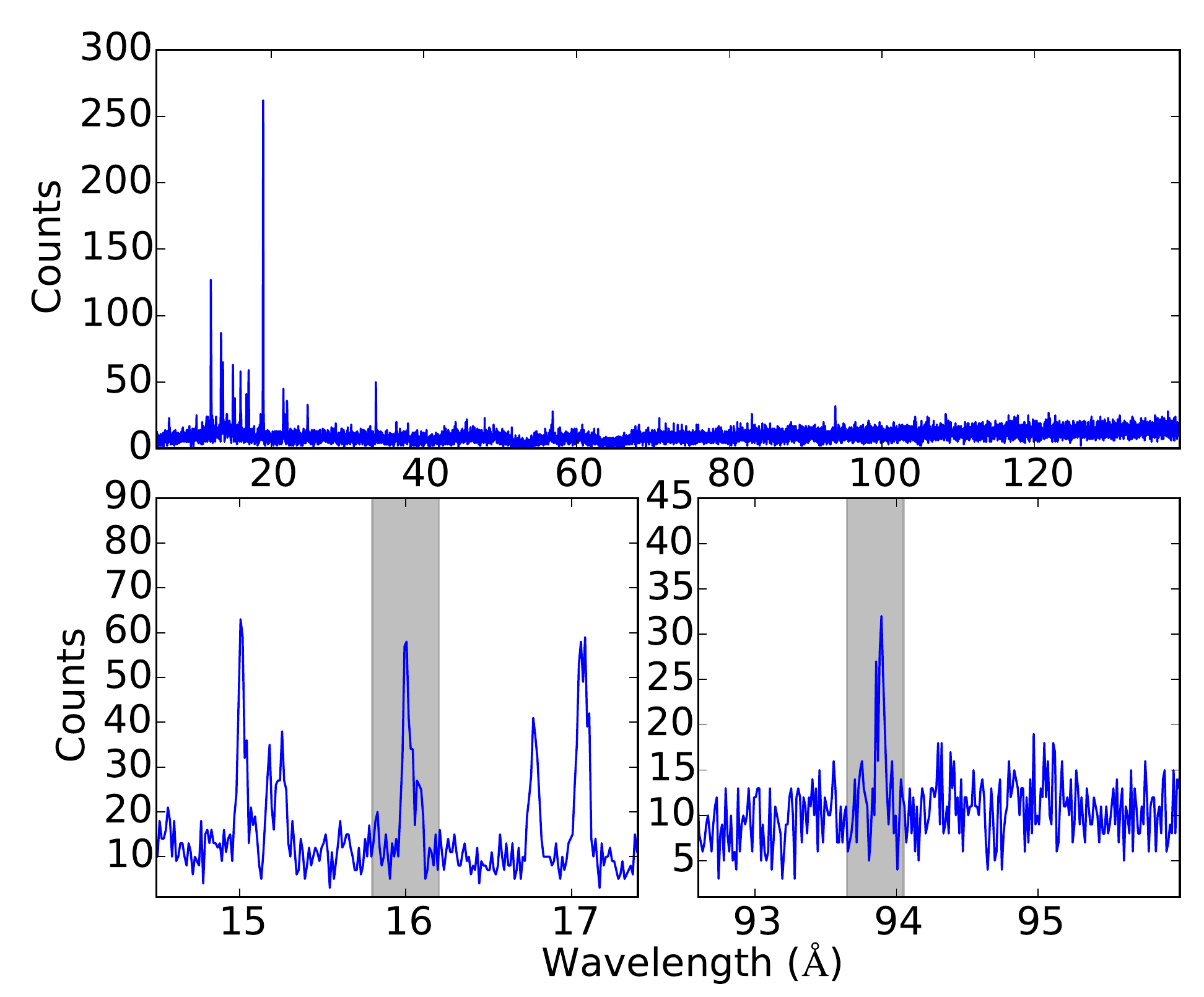} 
\includegraphics[width=2.6in]{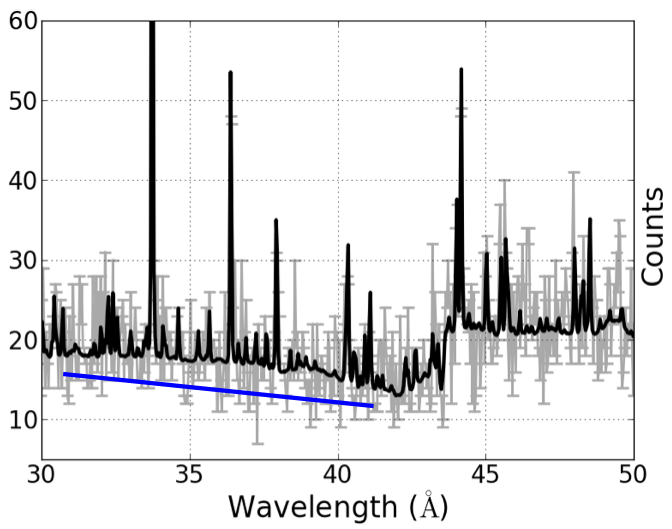} 
 \caption{\textbf{Left}: \emph{Chandra} LETGS spectrum of AU~Mic. As an example, two Fe~{\sc xviii} lines
 that were used to derive the absorbing column density are highlighted. \textbf{Right}: The carbon K-edge.
 The edge at 43\,\AA{} is clearly visibile, but can be entirely explained by carbon within the filter in
 front of the X-ray detector. Additional carbon within the AU~Mic disk would lower the flux shortwards
 of 43\,\AA{} as indicated by the blue line.}
   \label{fig:au_mic}
\end{center}
\end{figure}

Thanks to its youth and distance, AU~Mic is a bright X-ray source (e.g., \cite{Linsky_2002}).
We obtained a \emph{Chandra} LETGS X-ray spectrum of AU~Mic
(see Fig.~\ref{fig:au_mic}, published in \cite{Schneider_2012}), which extends from a
\hbox{few \AA{}} up to $>100\,$\,\AA.
This wavelengths coverage allowed us to compare lines of the same element and ionization stage
formed at largely different wavelengths. The line ratios are given  mainly by atomic 
physics and almost independent of the plasma temperature. Thus, these line ratios can be used to
measure the total absorption towards AU~Mic. Specifically from
our analysis of abundant Fe~lines, we limit the gas column density to 
$N_H<10^{19}\,$cm$^{-2}$ ($1\sigma$ conf.). Furthermore, this spectrum allowed us to directly
constrain the amount of carbon in gas and small grains ($\lesssim0.5\,\mu$m) from the depth
of the carbon K-edge. 
Figure~\ref{fig:au_mic} (right)
shows a zoom into the wavelengths region around the C-K edge. The edge-like  structure is clearly visible.
However, its depth can be entirely explained by carbon in the detector itself.
We derive an upper limit on the circumstellar carbon content of $N_C<10^{18}\,$cm$^{-2}$.

Comparing our upper limits with published dust masses, we find that the disk must contain more
dust than gas. {The measurements of \cite{Fitzgerald_2007} suggest a  dust mass 
around $0.01\,M_\oplus$ (\cite{France_2007}), which would correspond to a total dust column density of $7\times10^{-4}$\,g\,cm$^{-2}$ 
or $3\times10^{20}$\,cm$^{-2}$ for ISM abundances.}
That we detect less absorption implies that the dust is locked in large grains that are opaque to X-rays.

\vspace*{-0.4cm}
\section{The protoplanetary disk around the CTTS AA~Tau}

\begin{figure}[t]
\begin{center}
\includegraphics[width=2.65in, height=2.in]{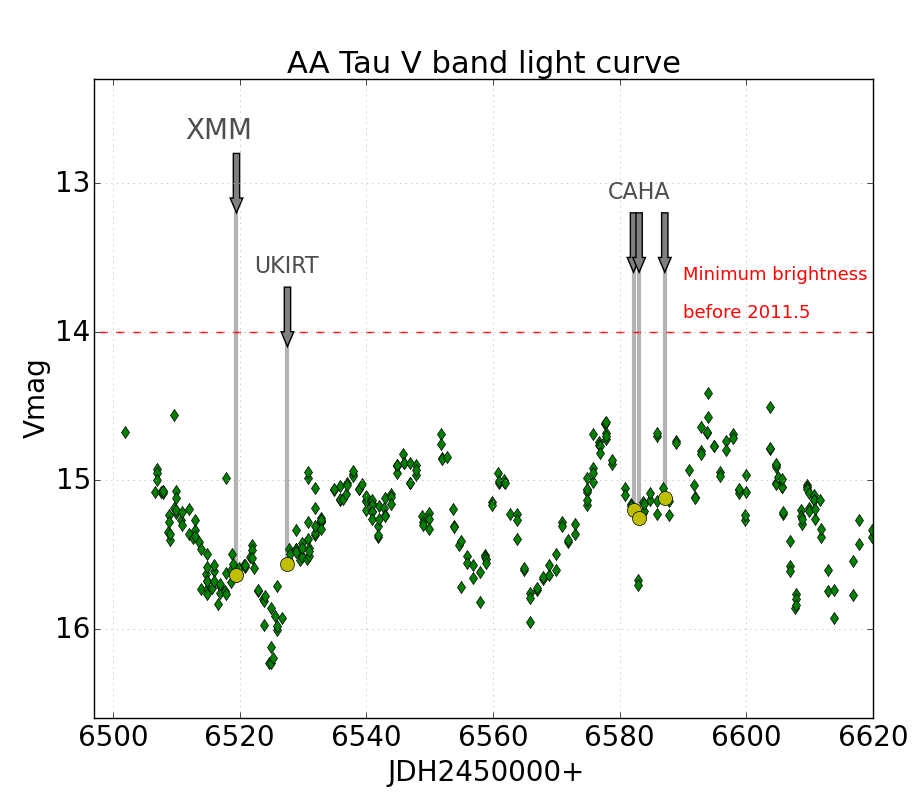} 
\includegraphics[width=2.6in, height=2.in]{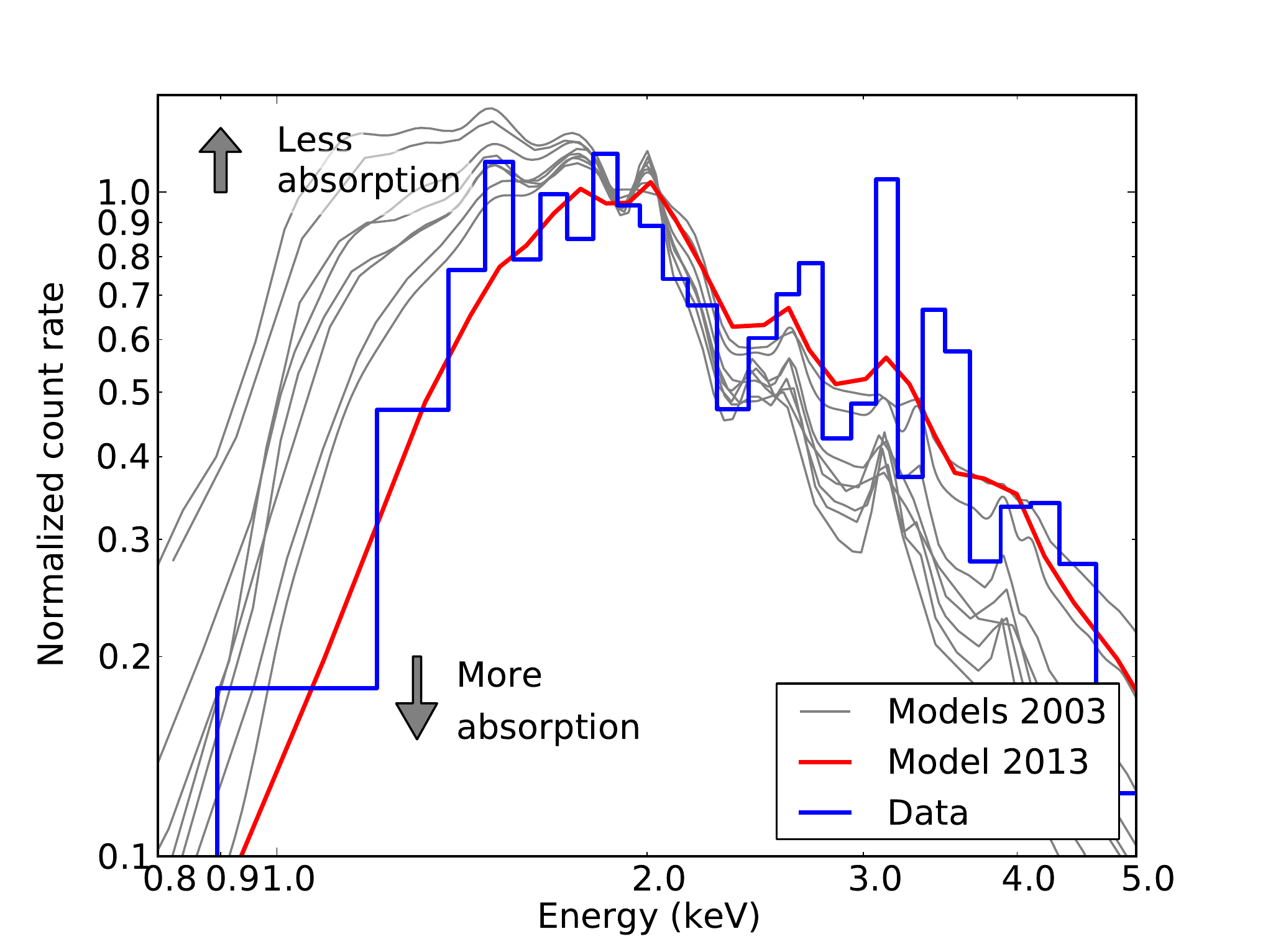} 
 \caption{\textbf{Left}: Optical light curve of AA~Tau around the X-ray and near-IR observations. 
 \textbf{Right}: XMM-Newton pn spectrum of AA~Tau with models normalized to the count rate at 
 $E_{phot}=2.0\,$keV. }
   \label{fig:aa_tau}
\end{center}
\end{figure}

The classical T~Tauri star (CTTS) AA~Tau is surrounded by a gaseous protoplanetary disk. Compared
to AU~Mic, AA~Tau is younger ($\sim2\,$Myrs) and its disk is not viewed edge-on, but inclined by 75$^\circ$
with respect to the plane of the sky. Thus, the disk is not in front of the
star for most of the time. The inner disk region, however, is warped and periodically eclipses the star. 
X-ray data showed that this inner region is gas-rich as measured by $N_H$/$A_V$ where $N_H$
is the equivalent hydrogen column density derived from low-resolution X-ray spectra 
and $A_V$ is the dust extinction (\cite{Schmitt_2007}, \cite{Grosso_2007}).

Towards the end of  2011, the system darkened significantly in the optical ($\Delta V \approx 2$, see 
Fig.~\ref{fig:aa_tau} left). \cite{Bouvier_2013}
suggest  that the extra absorption is caused by disk material at larger radii ($r\gtrsim10\,$AU), which 
rotated into view and now obscures the line-of-sight towards the central star. 
{ That the inner disk region is obscured from view is confirmed by our high-resolution 
HST FUV spectrum obtained during
the dim state, as the H$_2$ line width decreased compared
to the bright state's data published by \cite{France_2012}.} 
We obtained a new
X-ray observation with XMM-Newton during the dim state. It reveals that the increase in absorbing column density
is moderate (\hbox{$N_H^{extra}=0.5-1.0\times10^{22}$\,cm$^{-2}$}). 
Our contemporaneous UV to near-IR data characterize the dust extinction during the dim state.
Comparing dust and gas absorption, we find a gas-to-dust
ratio of $N_H/A_V=0.8\dots3.6\times10^{21}$\,cm$^{-2}$, which is within a factor of two of the ISM value 
(\cite{Schneider_2015a}). 

\vspace*{-0.4cm}
\section{RW Aur: Another CTTS during a dim state}
RW~Aur consists of two CTTSs; both of approximately solar mass.
{ The system underwent a major dimming event towards the end of 2014 ($\Delta V=$2 - 3\,mag).
This event resembles a previously observed half-year long dimming, which was attributed
to a tidal stream between both binary components
(\cite{Rodriguez_2013}, \cite{Dai_2015}). However, the new dimming might more
closely resemble the AA~Tau dimming with obscuration by the primary's disk itself as suggested by
infrared data (\cite{Shenavrin_2015}).}

\begin{figure}[t]
\begin{center}
\fbox{\includegraphics[width=2.5in]{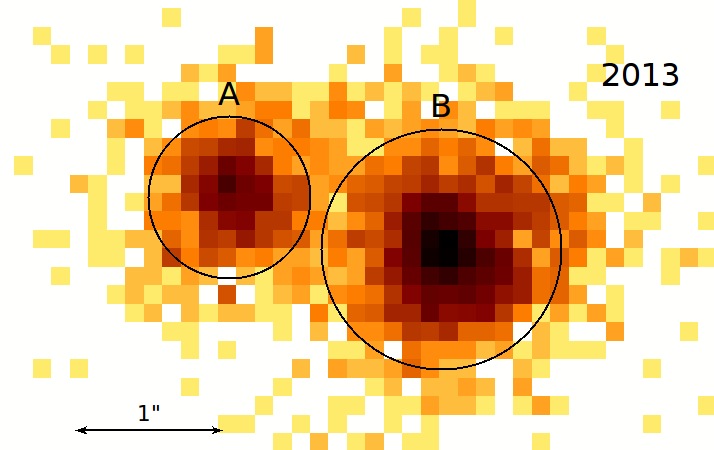}}
\fbox{\includegraphics[width=2.5in]{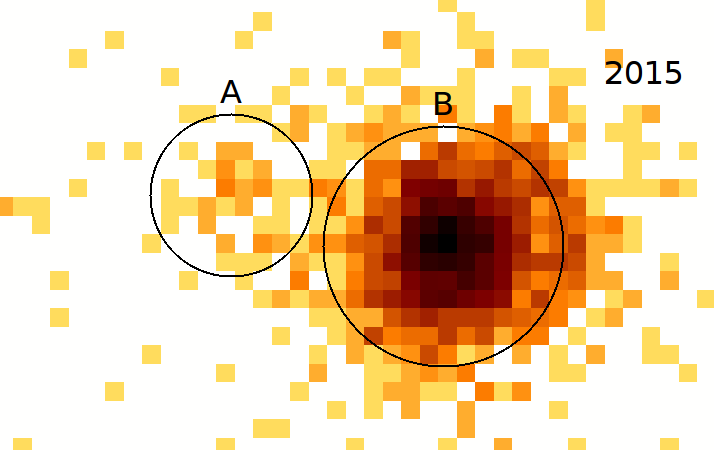}}
 \caption{\emph{Chandra} images of the RW~Aur system. \textbf{Left}: Bright state, \textbf{Right}: dim state with
 the tidal stream in front of the A component.
}
   \label{fig:rw_aur}
\end{center}
\end{figure}

We obtained a new \emph{Chandra} observation of the system during a dim state that started
end of 2014 (see Fig.~\ref{fig:rw_aur}). Comparing the dim state's X-ray properties with those observed previously by 
\cite{Skinner_2014} during the bright state, we
find an increase in absorbing column density of $N_H\approx2\times10^{22}$\,cm$^{-2}$. 
\cite{Antipin_20} find that the optical extinction appears rather gray, i.e., that the 
absorption is caused by grains with sizes greater than 1\,$\mu$m. Therefore, the
dust mass might differ from the value suggested by assuming $\Delta V = A_V$ and 
an ISM-like grain population. As an estimate,
we assume absorption by $1\,\mu$m sized grains and opacities from \cite{Draine_2003}).
With these assumptions, the gas-to-dust ratio of the extra absorber is below the ISM-value
when comparing the observed 2\,mag of dust extinction with the increase in X-ray
absorption (\cite{Schneider_2015b}).

\section{Conclusions}
The gas-to-dust ratio of circumstellar material is probed by the ratio between X-ray absorption
($N_H$) and optical/NIR extinction ($A_V$). We present three systems where we
applied this method. The debris disk system AU~Mic contains more grains than gas, the protoplanetary
disk around the CTTS AA~Tau has a gas-to-dust ratio compatible with the ISM at radii of a few AU, 
and the absorber in the young binary system RW~Aur has low gas-to-dust ratio and large grains.

%
%
%

\end{document}